\documentclass[10pt,a4paper]{article}
\usepackage[dvips]{color}
\usepackage{epsfig}
\usepackage{epstopdf}
\usepackage{amsmath}
\usepackage{graphicx}
\usepackage{authblk}
\usepackage{amssymb,amsmath}
\usepackage{amsmath}
\usepackage{tabularx}

\begin{document}
\title{\bf Can an interacting varying Chaplygin gas and tachyonic matter accelerate Universe?}
\author{{M. Khurshudyan$^{a,b,c}$\thanks{Email:khurshudyan@yandex.ru, khurshudyan@tusur.ru}
}\\
$^{a}${\small {\em International Laboratory for Theoretical Cosmology, Tomsk State University of Control Systems and Radioelectronics (TUSUR), 634050 Tomsk, Russia}}\\ 
$^{b}${\small {\em Research Division,Tomsk State Pedagogical University, 634061 Tomsk, Russia}}\\
$^{c}${\small {\em Institute of Physics, University of Zielona Gora, Prof. Z. Szafrana 4a, 65-516 Zielona Gora,
Poland}}\\
}\maketitle

\begin{abstract}
In this paper a possibility to accelerated the expansion of the large scale universe due to interacting varying Chaplygin gas of a specific form and a tachyonic matter is considered. A specific form of the potential and appropriate values of the parameters allow to obtain early tachyonic dark matter with $\omega_{T} \approx 0$ in recent Universe as well. On the other hand, related to the structure of the field equations describing the dynamics of the background we assumed the form of non-gravitational interaction to be given in advance. A sing changeable interaction between dark energy and dark matter is also included due to a huge interest towards it in recent literature. Possible future finite time singularities for cansidered models have been determined to finalize the paper. The study shows that in case of non-interacting model, the fate of the Universe will be either Type II~(“The Sudden Singularity), or Type IV~("Generalized Sudden Singularity") singularity. On the other hand, the consideration of $Q=3 b H \rho_{de}$ non-gravitational interaction in addition to Type II and Type IV singularities will induce also Type V~("$\omega$-singularity") singularity. Whilst in the model with $Q=3 b q H \rho_{de}$ non-gravitational interaction only Type IV singularity will be observed. Moreover, $2$ forms of the scale factor has been considered separately allowing analytically obtain the behavior of the cosmological parameters without specifying an explicit form of the potential for tachyonic matter. 
\end{abstract}

\section{\large{Introduction}}\label{sec:Int}

The explanation of recent observational data indicates about an existance of a dark side of unknown structure and origin in the large scale Universe. Various missions and observational programs allow systematic analysis of the Universe, however, still there are crucial open questions requiring phenomenological assumptions~\cite{Riess}. To explain the accelerated expansion of the Universe an impressive amount of hypothesis were proposed. For instance, if we consider general relativity to describe the background dynamics of the Universe, the desirable result could be achieved due to dark energy, which occupies about 70$\% $ of the energy of the large scale Universe. Dark energy is one of the components of existing darkness, while the second component is dark matter. It should be mentioned clearly, that dark energy and dark matter are effective quantities, and it is likely, that both of them are multicomponent. Therefore, there is a huge effort towards various parameterizations of the darkness to see which combination is the best according to the observational data and existing symmetries. There are various intriguing developments towards to dark energy problem. The simplest model of dark energy is cosmological constant $\omega_{\Lambda}=-1$, which has been introduced by Einstein for a different reason. On the other hand, cosmological constant creates two very serious problems. A possible way to alleviate cosmological constant problems an idea of dynamical dark energy has been suggested - quintessence, phantom, quintom~(scalar field models of dark energy), holographic dark energy and ghost dark energy~(to mention a few). On the other hand, dark fluids including viscous fluids also can be used~\cite{Steinhardt},~\cite{Guo},~\cite{Nojiri},~\cite{Hao}. 

Recent increasing interest towards different forms of non - gravitational interaction is due to observational data not presenting evidence against this idea and improving theoretical results on background and perturbative levels. On the other hand, we do not have any known symmetry which prevents or suppresses a non - minimal coupling between dark energy and dark matter. The question related to the microscopic description of non-gravitational interactions is still open and requaries a systematic and careful comparison of the results with observational data to come to the final conclusion~(see for instance~\cite{Hao}). 

In recent literature, there are two alternative approaches developed giving desirable solutions to the accelerated expanding problems. One of these alternatives is a modification of general relativity. There is a huge excellent discussion on this topic in recent scientific literature including astrophysical applications as well~\cite{MGR}. On the other hand, the third possibility is related to the creation of cold dark matter, which gives sufficient negative creation pressure to accelerate the Universe. It is obvious that such scenario is free from all problems which are related to dark energy mentioned above. Recently several interesting scenarios have been considered related to this possibility~\cite{PCR}. It should be mentioned that all possibilities to explain the dynamics of the large scale Universe is mainly operated on the phenomenological level. Presented model of this paper is also a phenomenological model, where a modified varying Chaplygin gas~(like other models of dark fluid Chaplygin gas has various modifications)~\cite{Zong}~(and references therein) 
\begin{equation}\label{eq:ChGas}
P_{c}=-\frac{A_{0}a^{-m}}{\rho^{\alpha}_{c}},
\end{equation}
where parameter $A$ depends on scale factor $a$ with $A_{0}>0$ and $m$ being constants, is applied to explain the accelerated expansion of the low redshift Universe. Moreover, the second component of the dynamical dark side is assumed to be tachyonic field with a relativistic Lagrangian~\cite{Sen}
\begin{equation}\label{eq:TF}
L=-V(\phi)\sqrt{1-\partial_{\mu}\phi\partial^{\mu}\phi},
\end{equation}
providing the following energy density
\begin{equation}\label{eq:TFdensity}
\rho_{T} = \frac{V(\phi)}{\sqrt{1-\dot{\phi}^{2}}}.
\end{equation}
and pressure
\begin{equation}\label{eq:TFpressure}
P_{T} = -V(\phi)\sqrt{1-\dot{\phi}^{2}}.
\end{equation}

For considered case equations describing the dynamics of a flat low redshift Universe with FRW metric and interacting dark components read as
\begin{equation}\label{eq: Fridmman vlambda}
H^{2}=\frac{\dot{a}^{2}}{a^{2}}=\frac{\rho}{3},
\end{equation}
\begin{equation}\label{eq:Freidmann2}
\frac{\ddot{a}}{a}= - \frac{1}{6}(\rho + 3 P),
\end{equation}
\begin{equation}\label{eq:inteqm}
\dot{\rho}_{c}+3H(\rho_{c}+P_{c}) = -Q,
\end{equation}
and
\begin{equation}\label{eq:inteqG}
\dot{\rho}_{T}+3H(\rho_{T}+P_{T} )= Q.
\end{equation}
Last two equations describe non - gravitational interaction providing a transition from dark energy to dark matter.

In this paper an attempt to explore the acceleration expansion of the large scale Universe involving varying Chaplygin gas of a specific form with tachyonic matter is discussed. In the first part of the paper, we will study the model with a specific form of the potential $V(\phi)$ for tachyonic field giving early tachyonic cold dark matter with EoS equal $0$. Moreover, we will demonstrate that such dark matter has a transition from early cold dark matter to dark energy and for low redshifts again a transition to cold dark matter will be observed. On the other hand, the impact on the background dynamics of two specific forms of non-gravitational interactions 
\begin{equation}\label{eq:Q1}
Q=3 b H \rho_{de},
\end{equation}
and
\begin{equation}\label{eq:Q2}
Q= 3 b q H \rho_{de},
\end{equation}
where $b$ is a constant, $\rho_{de}$ it is the energy density of dark energy, while $q$ it is the deceleration parameter, will be studied.  

In the second part of the paper, two different forms of the scale factor to describe the low redshift Universe 
\begin{equation}\label{eq:a1}
a(t)=t^{n}, 
\end{equation}
\begin{equation}
a(t)=a_{0}t^{m}\exp[\alpha t],
\end{equation}
where $n$, $m$, $a_{0}$ and $\alpha$ are constants, will be considered. The last form of the scale factor describes the Universe in quasi - exponential phase~\cite{Murli}. Due to the structure of the field equations a given form of the scale factor allows to study the models without specifying the form of the potential for the tachyonic matter. Moreover, when the forms of the scale factor are given, the potential and field can be recovered numerically when the dynamics of the energy densities of dark energy and dark matter are known. $Om$ analysis with a clasification of future finite time singularities for considered models will be discussed as well~\cite{Om},~\cite{Sing},~\cite{Sing1},~\cite{Sing2}.\\

The paper is organized as follows: After Introduction, in section~\ref{sec:PF} the models with a specific form of $V(\phi)$ potential and various forms of non-gravitational interactions between dark energy and dark matter are analysed. On the other hand, in section~\ref{sec:OMCS} $Om$ analysis and the classification of future finite time singularities for considered models will be presented. Particular attention is on the influance of non-gravitational interaction on the type of the singularity. In section~\ref{sec:SF} for two forms of the scale factor andinteraction term we found analytical forms of the energy densities of dark energy and dark matter, allowing to recover the scalar field and potential of interacting tachyonic dark matter numerically. Finally, discussion on obtained results are summarised in section~\ref{sec:Disc}.

\section{Tachyonic matter with specific $V(\phi)$ potential}\label{sec:PF}

\begin{figure}[h!]
 \begin{center}$
 \begin{array}{cccc}
\includegraphics[width=80 mm]{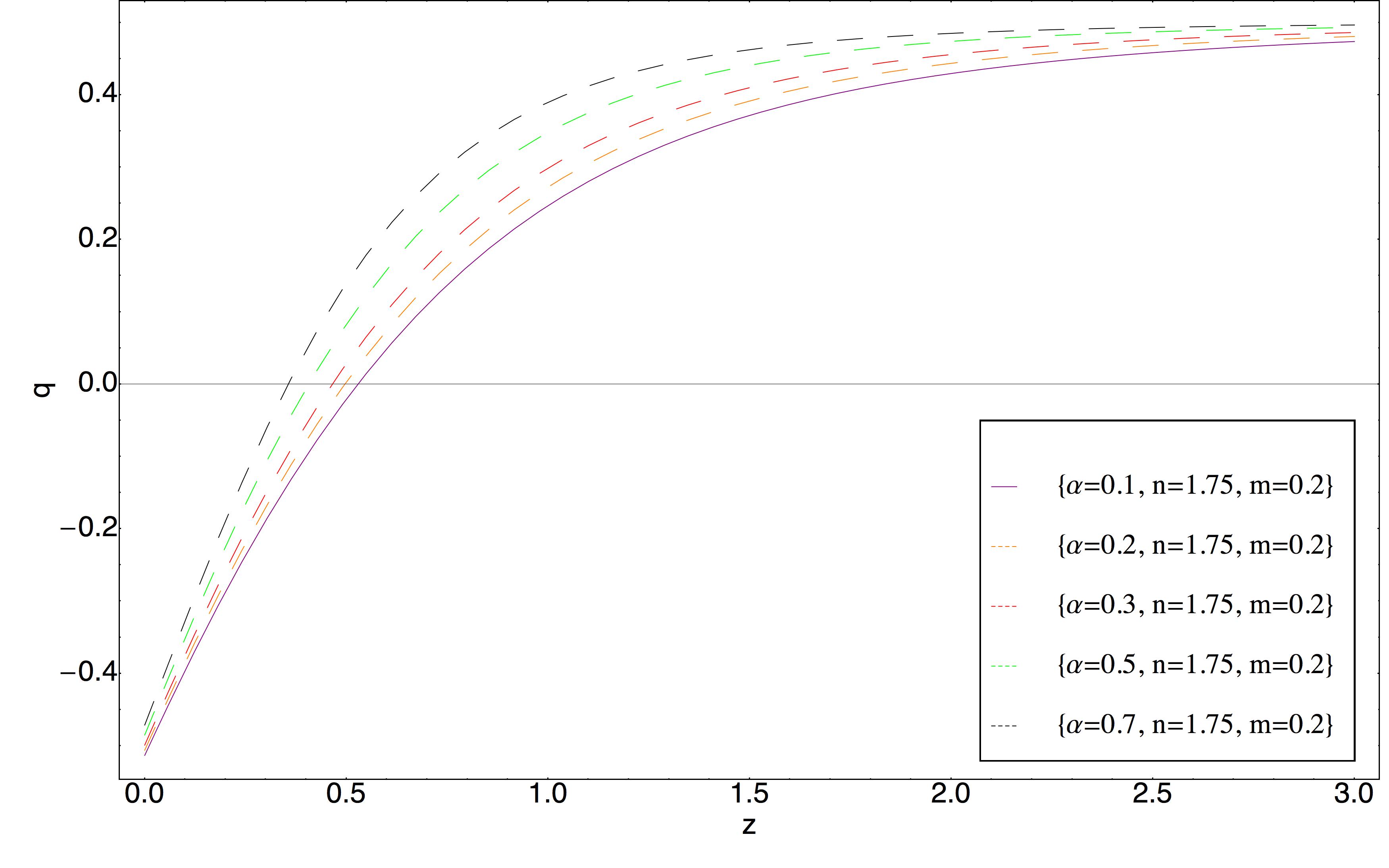}  &
\includegraphics[width=80 mm]{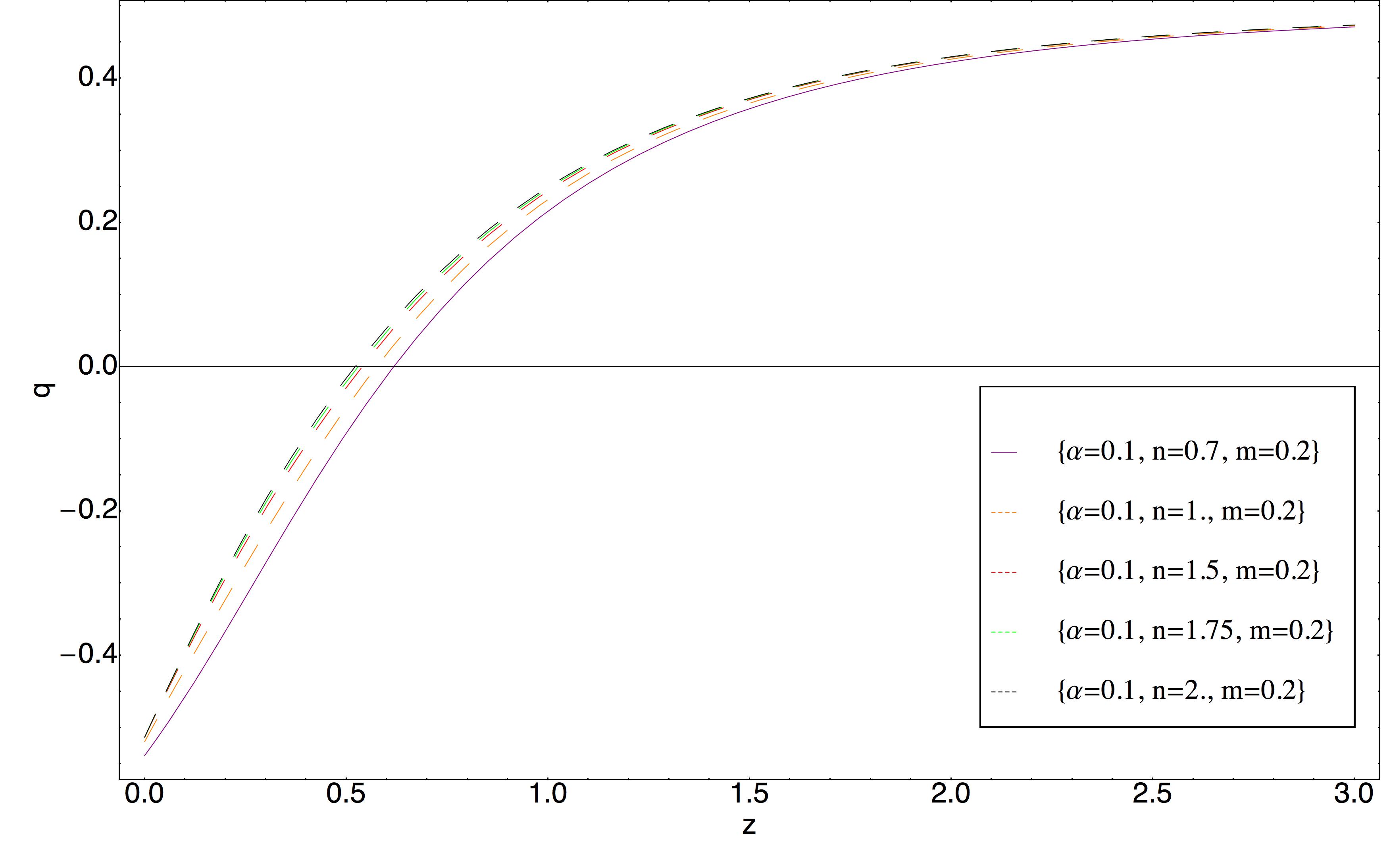}  \\
\includegraphics[width=80 mm]{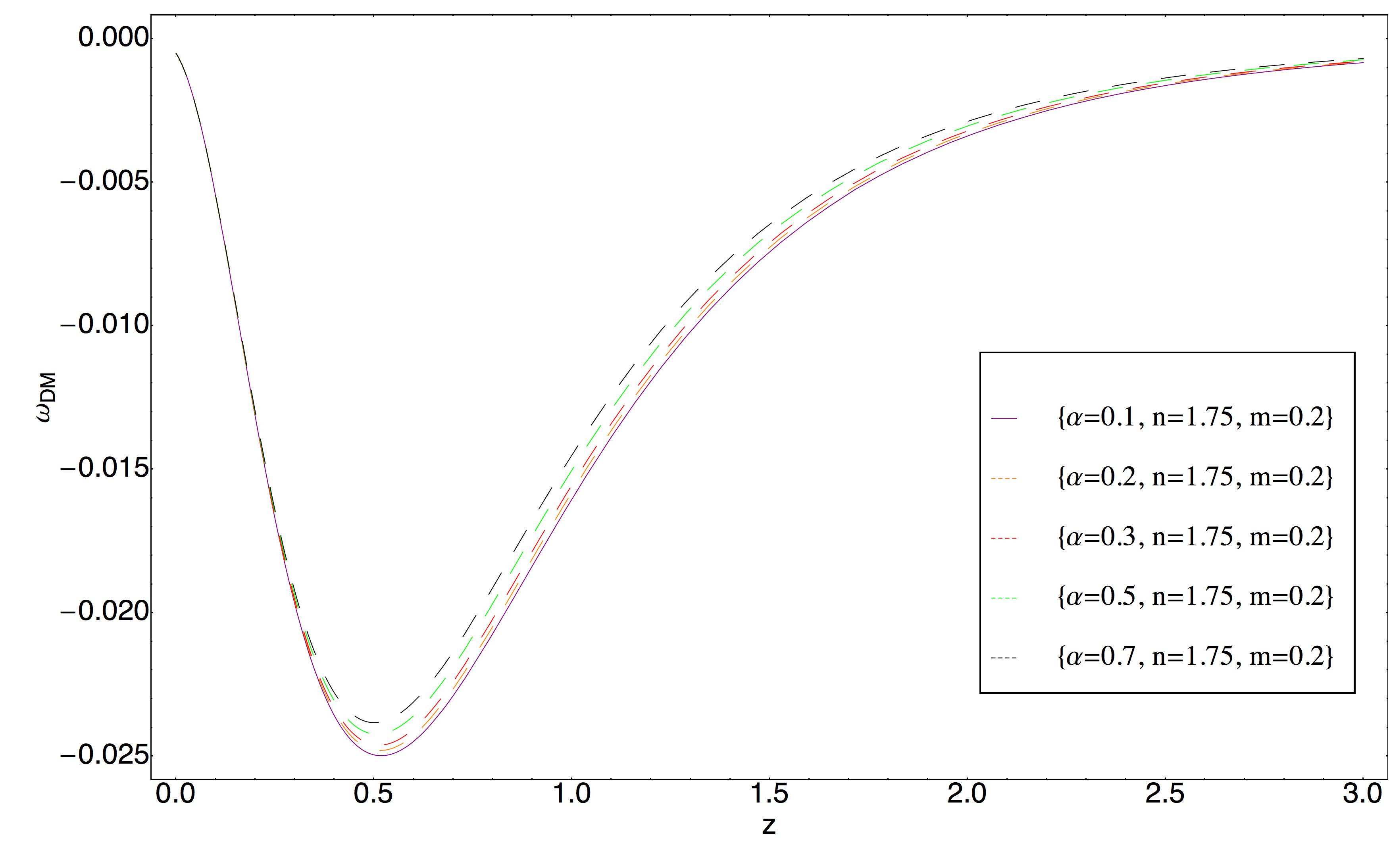}  &
\includegraphics[width=80 mm]{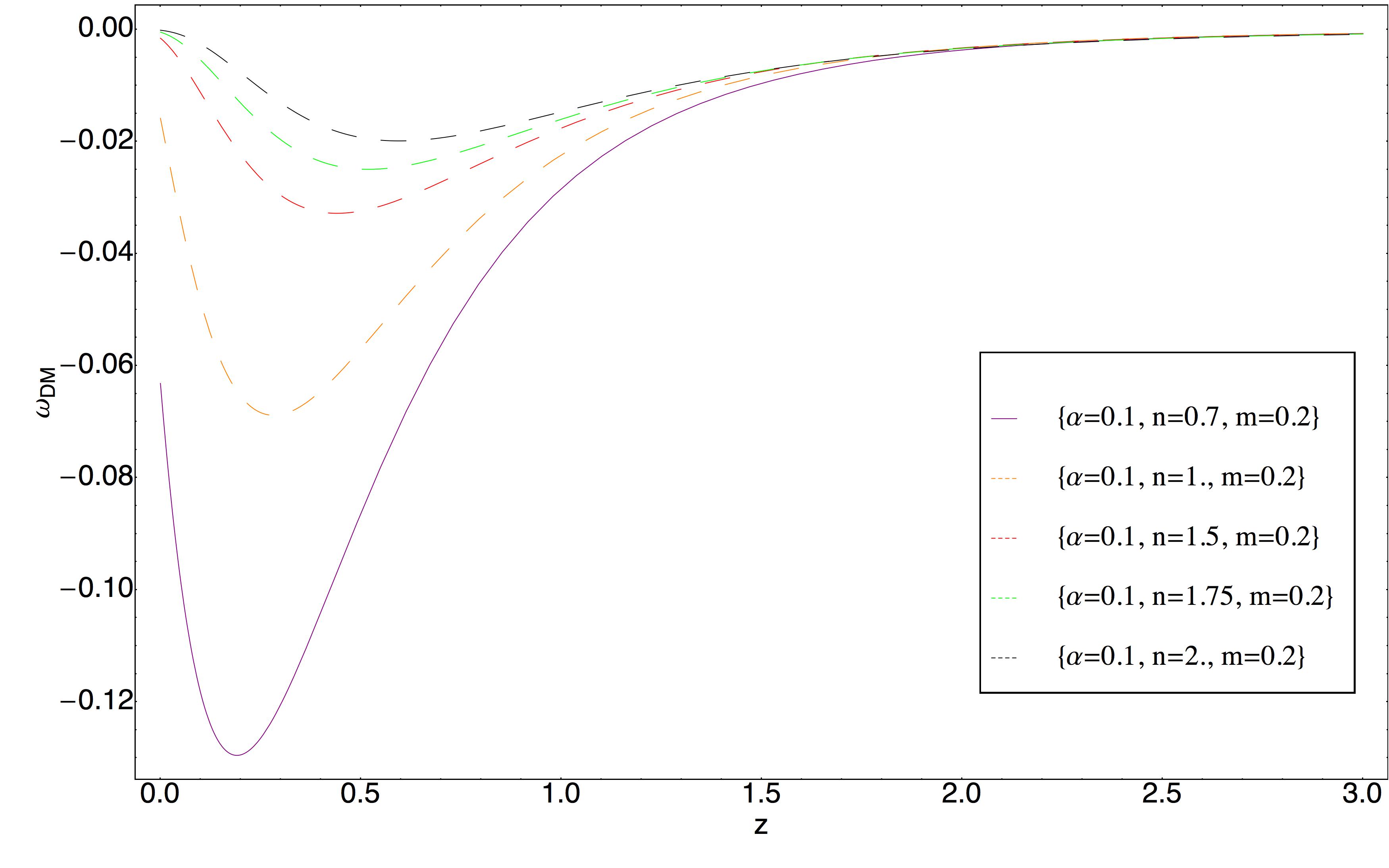}  \\
 \end{array}$
 \end{center}
\caption{The graphical behavior of the deceleration parameter $q$ and $\omega_{DM}$ of the tachyonic dark matter with $V(\phi) = V_{0} \phi^{n}$ against redshift $z$. Non-interacting model.}
 \label{fig:Fig1}
\end{figure}

We concentrate our attention only on such ranges of the parameters of the models when the tachyonic field behaves  as a cold dark matter in the early and recent large scale Universe. Thanks to the second component which behaves as the dark energy one can obtain a low redshift Universe in good agreement with recent observational data. The non - interacting model is the minimal model of this paper and some results of the study are presented in Fig.(\ref{fig:Fig1}). In particular, Fig.~(\ref{fig:Fig1}) represents the graphical behavior of the deceleration parameter $q$ and EoS parameter $\omega_{DM}$ of the tachyonic field with $V(\phi) = V_{0} \phi^{n}$ potential where $V_{0}$ and $n$ are constants. The graphical behavior of $\omega_{DM}$ shows that with appropriate values of the model parameters we will have tachyonic cold dark matter in the early and recent Universe~(the bottom panel of Fig.~(\ref{fig:Fig1})). On the other hand, from Fig.~(\ref{fig:Fig1}), one sees that an increase of $\alpha$ parameter describing varying Chaplygin gas gives a decrease in transition redshift and increases the present day value of the deceleration parameter $q$. On the other hand, the present day value of $\omega_{DM}$ does not affected strongly by the change of $\alpha$ parameter. Moreover, the analysis shows that a significant impact rised due to the parameter $\alpha$ on the behavior of EoS of the tachyonic field will be observed for $z \in [0.4,2]$ redshifts. 

\begin{figure}[h!]
 \begin{center}$
 \begin{array}{cccc}
\includegraphics[width=80 mm]{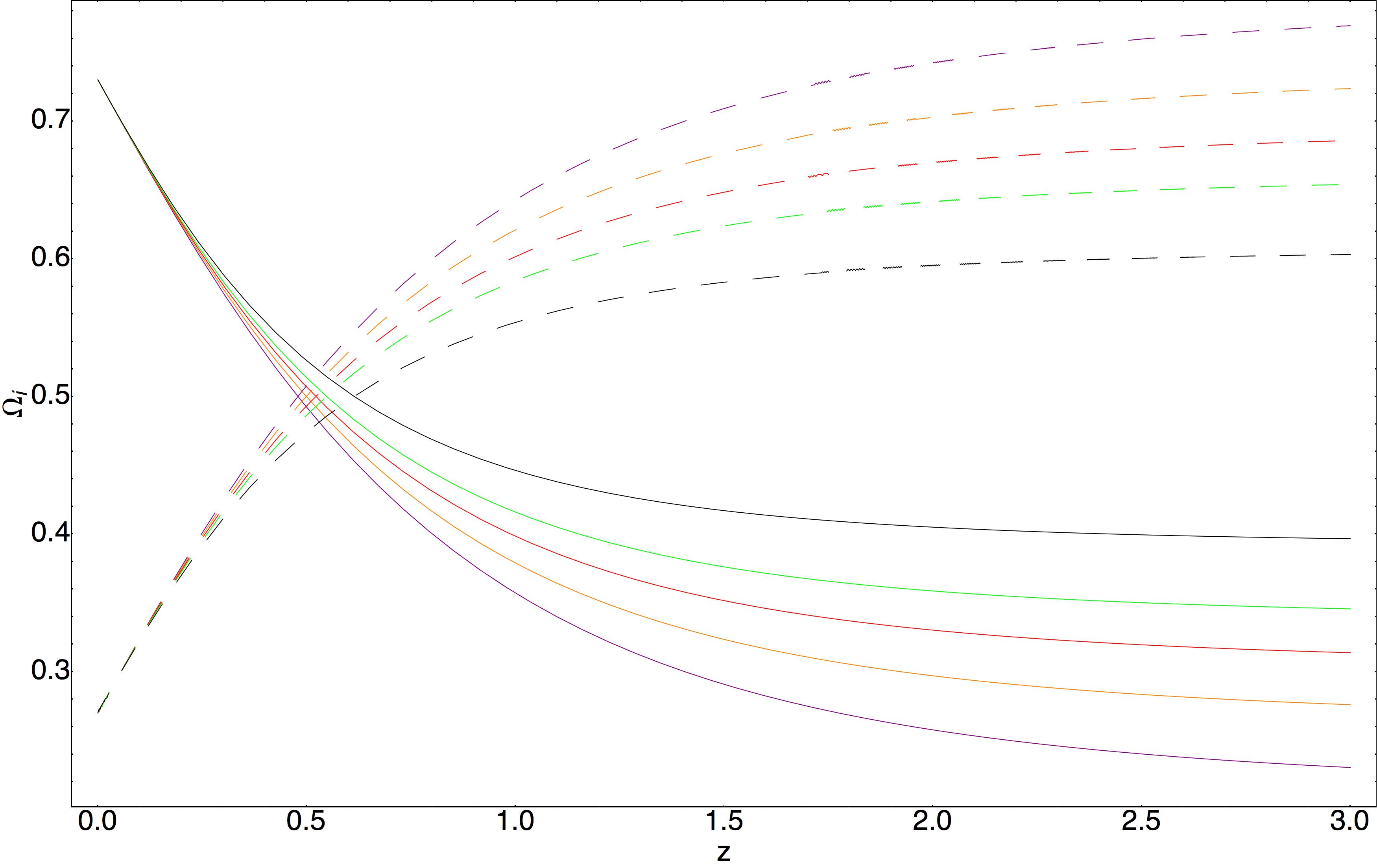}  &
\includegraphics[width=80 mm]{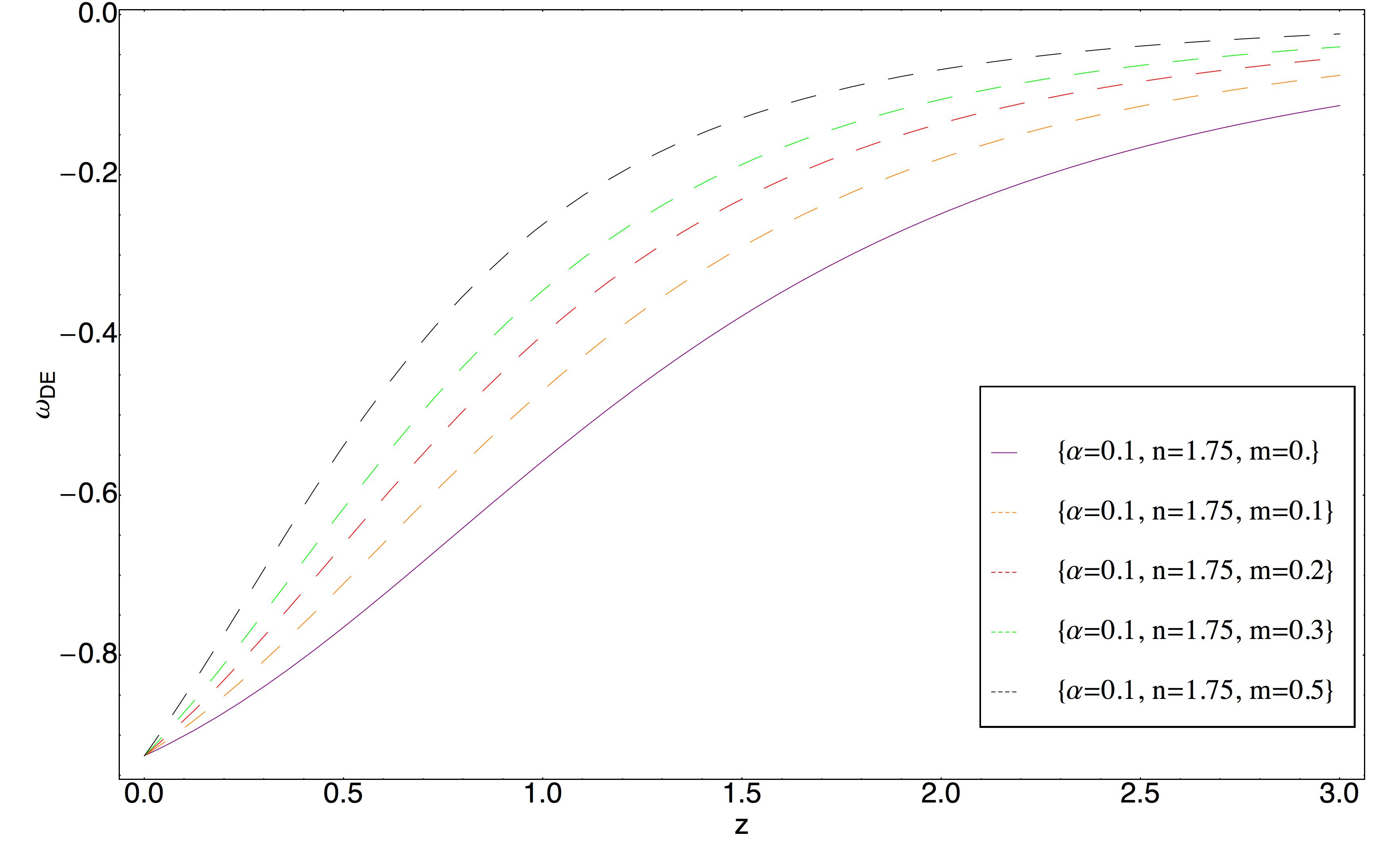}  \\
 \end{array}$
 \end{center}
\caption{The graphical behavior of $\Omega_{DE}$, $\Omega_{DM}$ and $\omega_{DE}$ of the varying Chaplygin gas given by Eq.~(\ref{eq:ChGas}) against redshift $z$. The potential of the tachyonic dark matter is given by $V(\phi) = V_{0} \phi^{n}$. Non-interacting model.}
 \label{fig:Fig2}
\end{figure}

On the other hand, the Fig.(\ref{fig:Fig1}) demonstrates the impact of the $n$ parameter on the behavior of the deceleration parameter and $\omega_{DM}$. One sees, that an increase of $n$ will decrease the transition redshift. Moreover, it can be seen that an increase of $n$ has a significant impact on the dynamics of $\omega_{DM}$ for the redshift $z < 1.0$ providing transition from the dark energy to the cold dark matter with $\omega_{DM} \approx 0$. The analysis shows that the parameter $n$ plays the central role in the transition from dark energy to cold dark matter in the recent Universe. On the other hand, Fig.~(\ref{fig:Fig2}) represents the behavior of $\Omega_{DE}$, $\Omega_{DM}$ and $\omega_{DE}$ for different values of the parameter $m$ demonstrating that the model is free from the cosmological coincidence problem. Moreover, the evolution of $\omega_{DE}$ shows that in this specific scenario varying Chaplygin gas is a matter with $\omega_{DE} \approx 0$ and during the evolution of the Universe will transfer into a quintessence dark energy. One sees, that it is possible to parametrize the energy budget of the Universe using a varying Chaplygin gas and a tachyonic matter with a specific form of the potential to explain the accelerated expansion of the large scale universe free from the cosmological coincidence problem. Moreover, the considered tachyonic matter can replace the cold dark matter in the resent Universe. 

\begin{figure}[h!]
 \begin{center}$
 \begin{array}{cccc}
\includegraphics[width=80 mm]{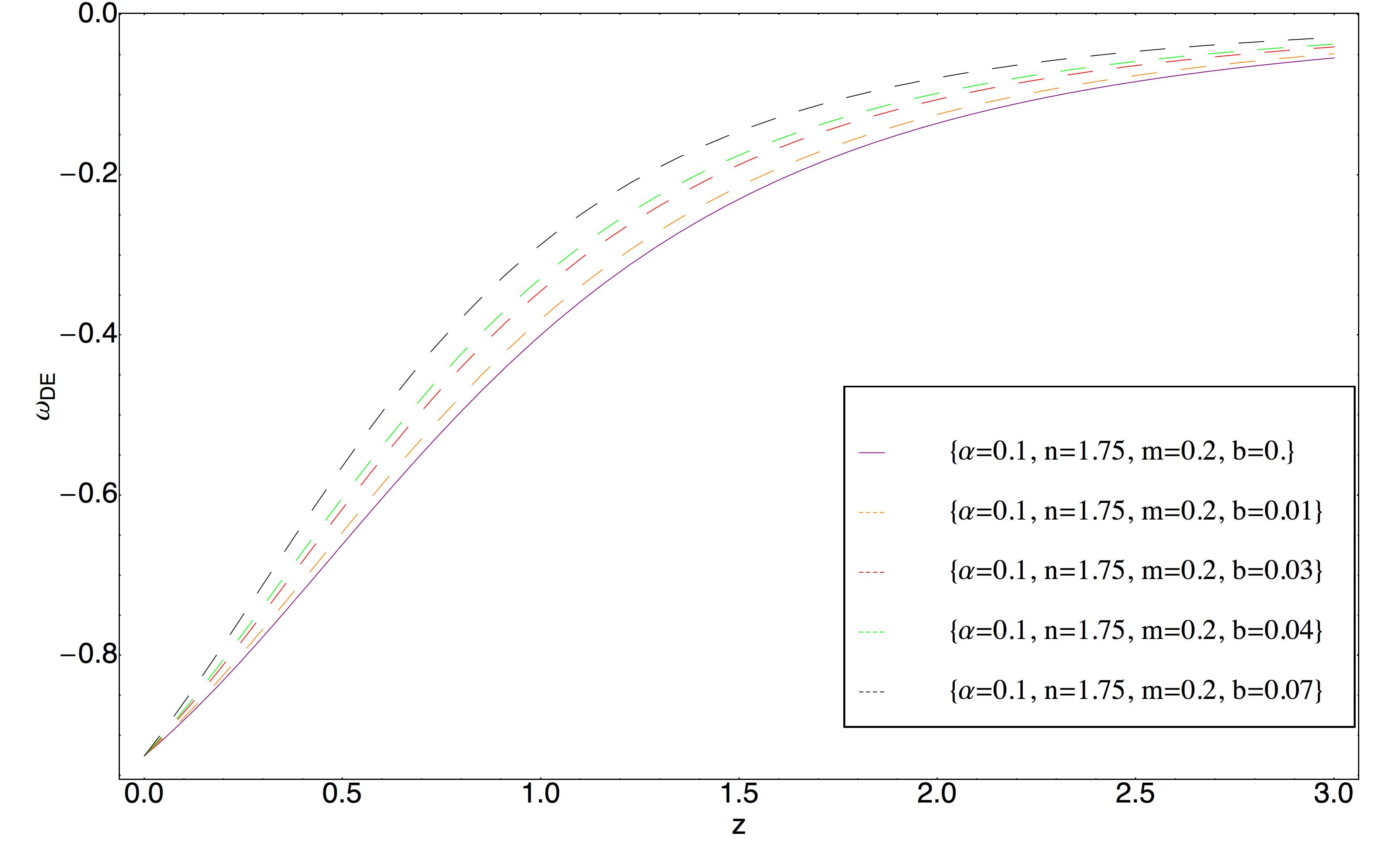}  &
\includegraphics[width=80 mm]{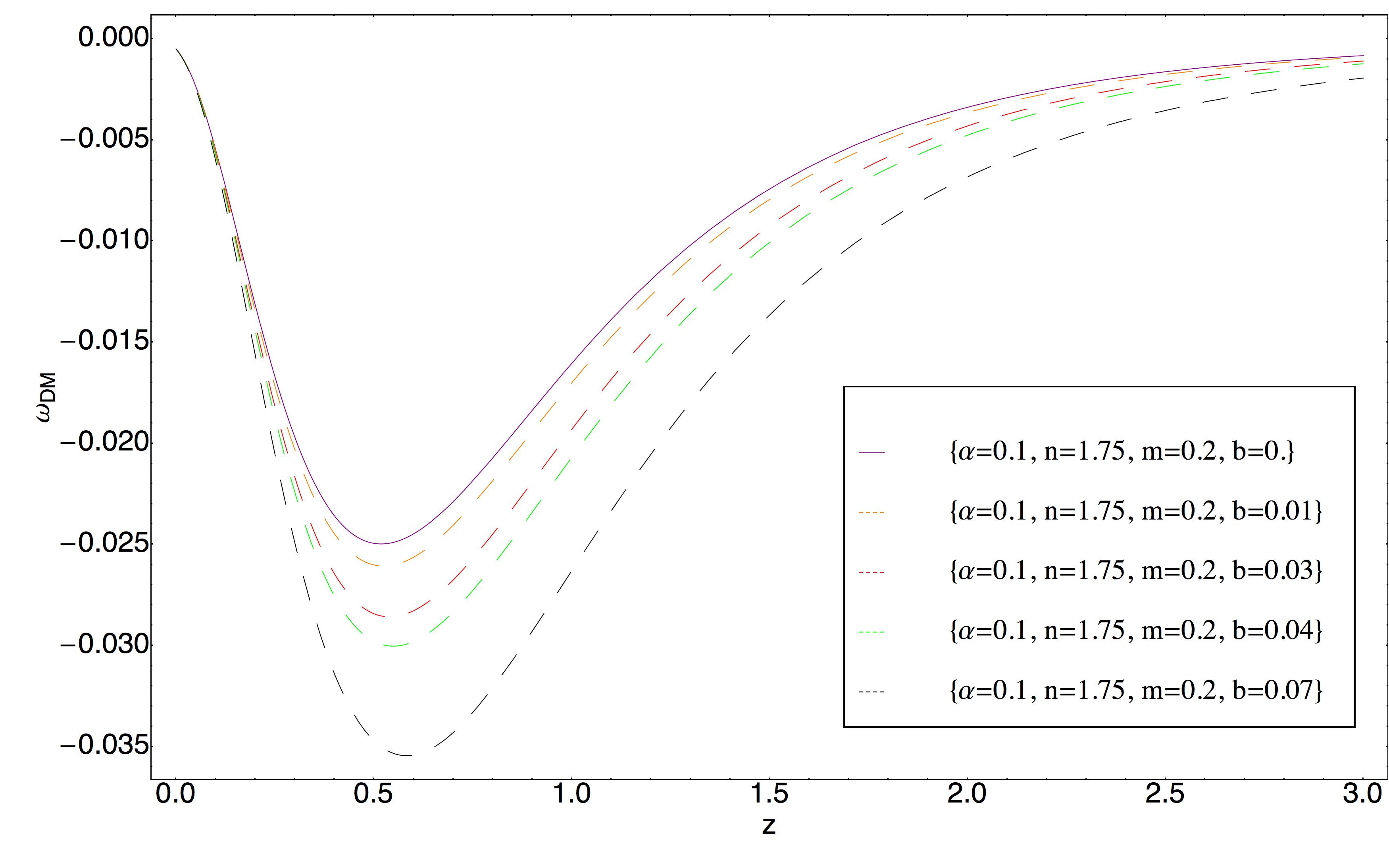} \\
\includegraphics[width=80 mm]{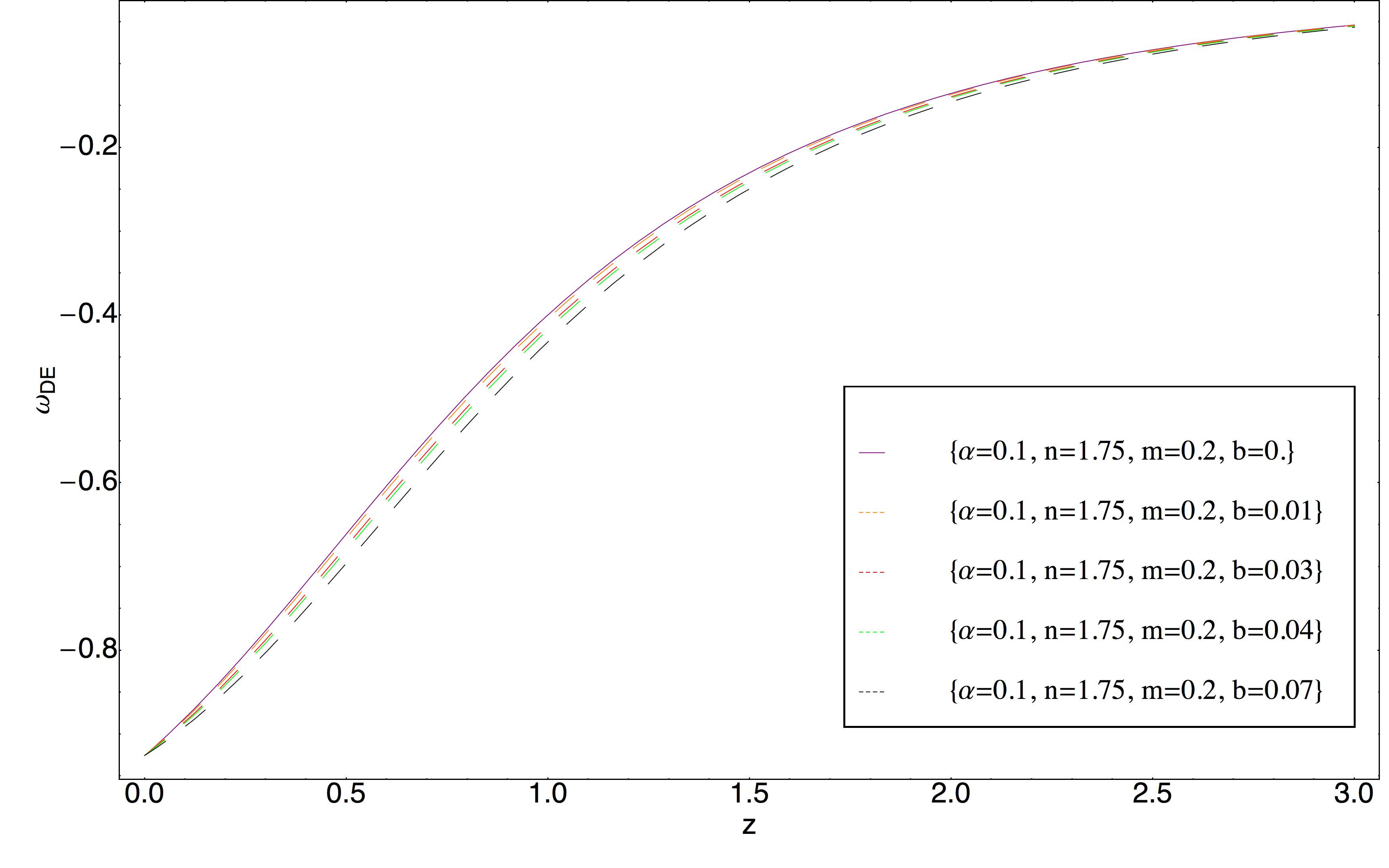}   &
\includegraphics[width=80 mm]{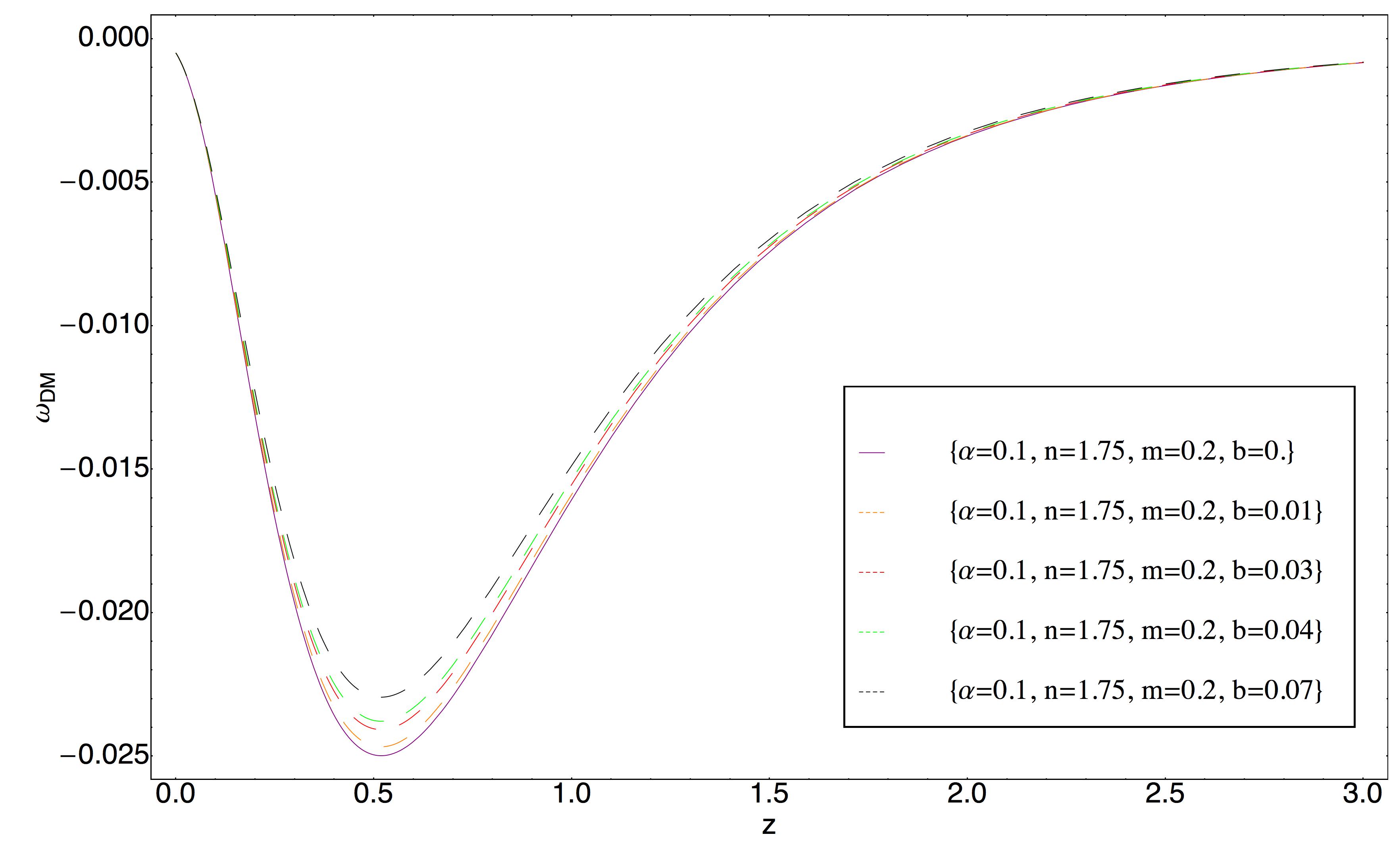}  \\
 \end{array}$
 \end{center}
\caption{The graphical behavior of $\omega_{DE}$ of the varying Chaplygin gas given by Eq.~(\ref{eq:ChGas}) and $\omega_{DM}$ of the tachyonic dark matter with $V(\phi) = V_{0} \phi^{n}$ against redshift $z$. The top panel represents the graphical behavior of $\omega_{DE}$ and $\omega_{DM}$ for the model with non-gravitational interaction given by Eq.~(\ref{eq:Q1}). The bottom panel represents the graphical behavior of $\omega_{DE}$ and $\omega_{DM}$ for the model with non-gravitational interaction given by Eq.~(\ref{eq:Q2}).}
 \label{fig:Fig3}
\end{figure}

To complete the study of the background dynamics two forms of non - gravitational interactions, Eq.~(\ref{eq:Q1}) and Eq.~(\ref{eq:Q2}), have been included into the model and the graphical behavior of EoS parameter for the varying Chaplygin gas and tachyonic matter is presented on Fig.~(\ref{fig:Fig3}). In particular the impact of non-gravitational interactions, Eq.~(\ref{eq:Q1}) and Eq.~(\ref{eq:Q2}), has been studied for the case with $\alpha = 0.1$, $n=1.75$ and $m = 0.2$, when non - interacting model has been desctribed by $z_{tr} \approx 0.55$ and $\omega_{DM} \approx 0$. In general, the study shows that in case of non-interacting model at $z=0$ tachyonic matter will be cold dark matter when $\alpha = 0.1$ and $m=0.2$, while $n \in [1.5, 2]$. We observed that the considered interactions will not affect the latter nature of the tachyonic matter described by $\omega_{DM}$. In particular, in case of non-gravitational interaction, Eq.~(\ref{eq:Q1}), we observed that for $z < 0.1$ considered interaction will not play a role on the dynamics of $\omega_{DM}$ and the maximal change for $\omega_{DM}$ compared with non-interacting case will be observed at $z \approx 0.56$. However, in case of the sign changeable interaction, Eq.~(\ref{eq:Q2}), an increase of the parameter $b$ will increase $\omega_{DM}$ for $z\in [0.2,1.4]$, while will decrease the $\omega_{DE}$. Moreover, for redshifts $z < 0.2$ and $z > 1.4$ the trace of non-gravitational interaction, Eq.~(\ref{eq:Q2}), can not be detected from the behavior of $\omega_{DM}$. On the other hand, sign fixed interaction, Eq.~(\ref{eq:Q1}), will decrease $\omega_{DM}$, while will increase the $\omega_{DE}$ of the varying Chaplygin gas. Another interesting fact is that considered forms of non-gravitational interactions do not affect on transition redshift and deceleration parameter. In recent literature there is fastly growing number of theoretical and phenomenological models explaining the accelerated expansion of the Universe and we need a control on these models. In particular, it can be done by a systematic and careful comparison of the results with observational data. However, this is a costly procedure and model independent analysis can be an alternative solution. There are several alternative model independent analysis and one of them is $Om$ analysis which is a null test for $\Lambda$CDM standard model. Such analysis also can be very useful to indicate differences between different states of the same model, when the study of a cosmological parameter does not allow clearly see them. The next section is devoted to $Om$ analysis and the classification of finite time future cosmological singularities.

\section{$Om$ analysis and cosmological singularities}\label{sec:OMCS}

We already mentioned that it is extremely important to control the large number of cosmological models devoted to the solution of the problems of accelerated expanding universe. In particular, we mentioned that it can be done by a model independent analysis and $Om$ is among them. We know that $H(z)$ is a model independent quantity and $Om$ for FRW universe is a combination of the Hubble rate $H(z)$ and the redshift, which makes it very attractive. $Om$ analysis has been used intensively in recent literature and it can be used to distinguish several dark energy models from $\Lambda$CDM. $Om$ is defined in the following way~\cite{Om}
\begin{equation}
Om = \frac{h^{2}(z) - 1}{(1+z)^{3} - 1},
\end{equation} 
where $h(z) = H(z)/H_{0}$, and with $Om = \Omega^{(0)}_{dm}$ it is a null test for the $\Lambda$CDM model. On the other hand, $Om > \Omega^{(0)}_{dm}$ and $Om < \Omega^{(0)}_{dm}$ stand for quintessence and phantom nature of the model, respectively. The defination of $Om$ shows that to determine the nature of the model the precise constraints on the matter density parameter as well as $H_{0}$ i.e. the value of the Hubble parameter at $z=0$, are not necessary to be known.

In general, intuitively it is clear, that if one of the cosmological parameters blows up, then a singularity should be formed and according to Ref.~\cite{Sing} the following types of singularities appear~(first time appeared classification):
\begin{itemize}
\item Type I~("The BIg Rip Singularity"): If the singularity occurs at time instance $t = t_{s}$, then the scale factor $a$, the effective energy density $\rho_{eff}$ and the pressure $P_{eff}$, diverge as $t \to t_{s}$, that is, $a \to \infty$, $\rho_{eff} \to \infty$ and $|P_{eff}| \to \infty$. This case yields incomplete null and time-likegeodesics,
 
\item Type II~(“The Sudden Singularity): In this case, the scale factor $a$ and the total effective energy density $\rho_{eff}$ are finite, but the effective pressure $P_{eff}$ diverges as $t \to t_{s}$, that is, $a \to  a_{s}<\infty$, $\rho_{eff} \to \rho_{s}<\infty$ and $|P_{eff}| \to \infty$. In this case the geodesics are complete and observers are not necessarily crushed~(weak singularity),

\item Type III~("The Big Freeze Singularity"): In this case, only the scale factor is finite, and the effective pressure and effective density diverge at $t \to t_{s}$, that is, $a \to  a_{s}<\infty$, $\rho_{eff} \to \infty$ and $|P_{eff}| \to \infty$. These can be either weak or strong singularity, which are geodesically complete solutions.  

\item Type IV~("Generalized Sudden Singularity"): In this case, the scale factor, the effective pressure and the effective density are finite at $t \to t_{s}$. On the other hand, the Hubble rate and its first derivative are also finite, but the higher derivatives of the Hubble rate diverge at $t \to t_{s}$. In this case there occur weak singularity and the geodesics are complete.

\end{itemize}

\begin{figure}[h!]
 \begin{center}$
 \begin{array}{cccc}
\includegraphics[width=80 mm]{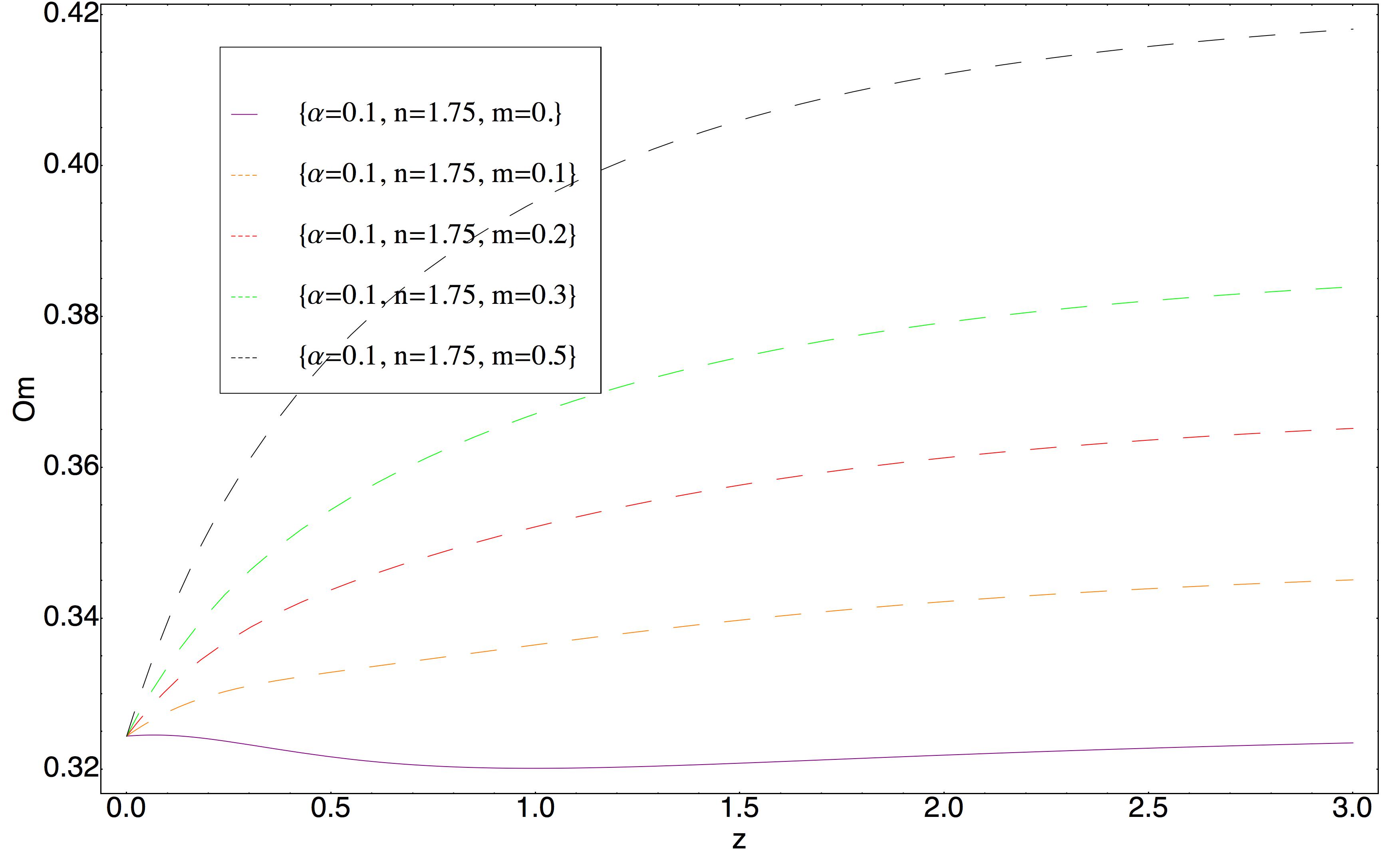}  &
\includegraphics[width=80 mm]{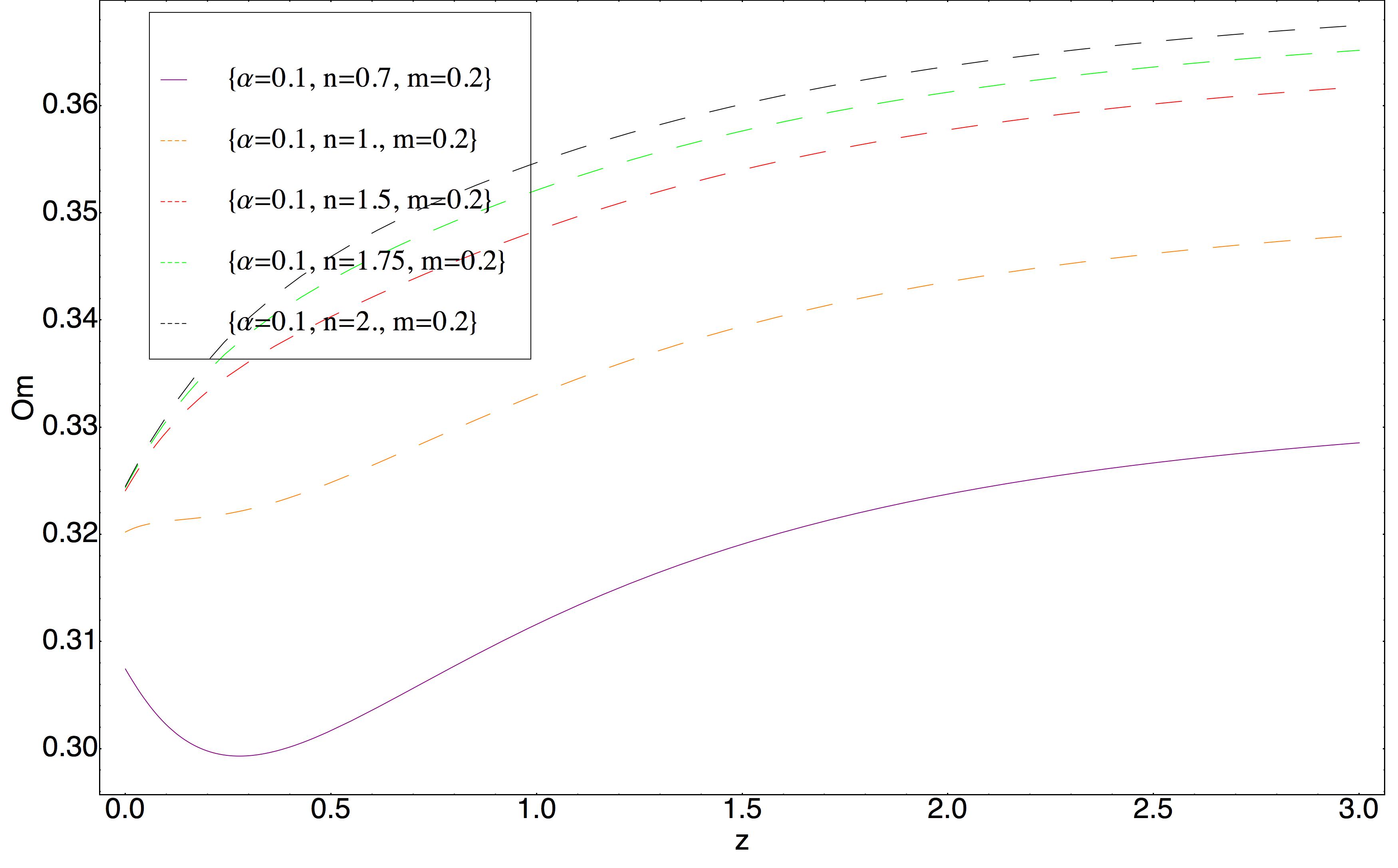}  \\
 \end{array}$
 \end{center}
\caption{The graphical behavior of $Om$ parameter against redshift $z$. Non-interacting model.}
 \label{fig:Fig4}
\end{figure}

\begin{figure}[h!]
 \begin{center}$
 \begin{array}{cccc}
\includegraphics[width=80 mm]{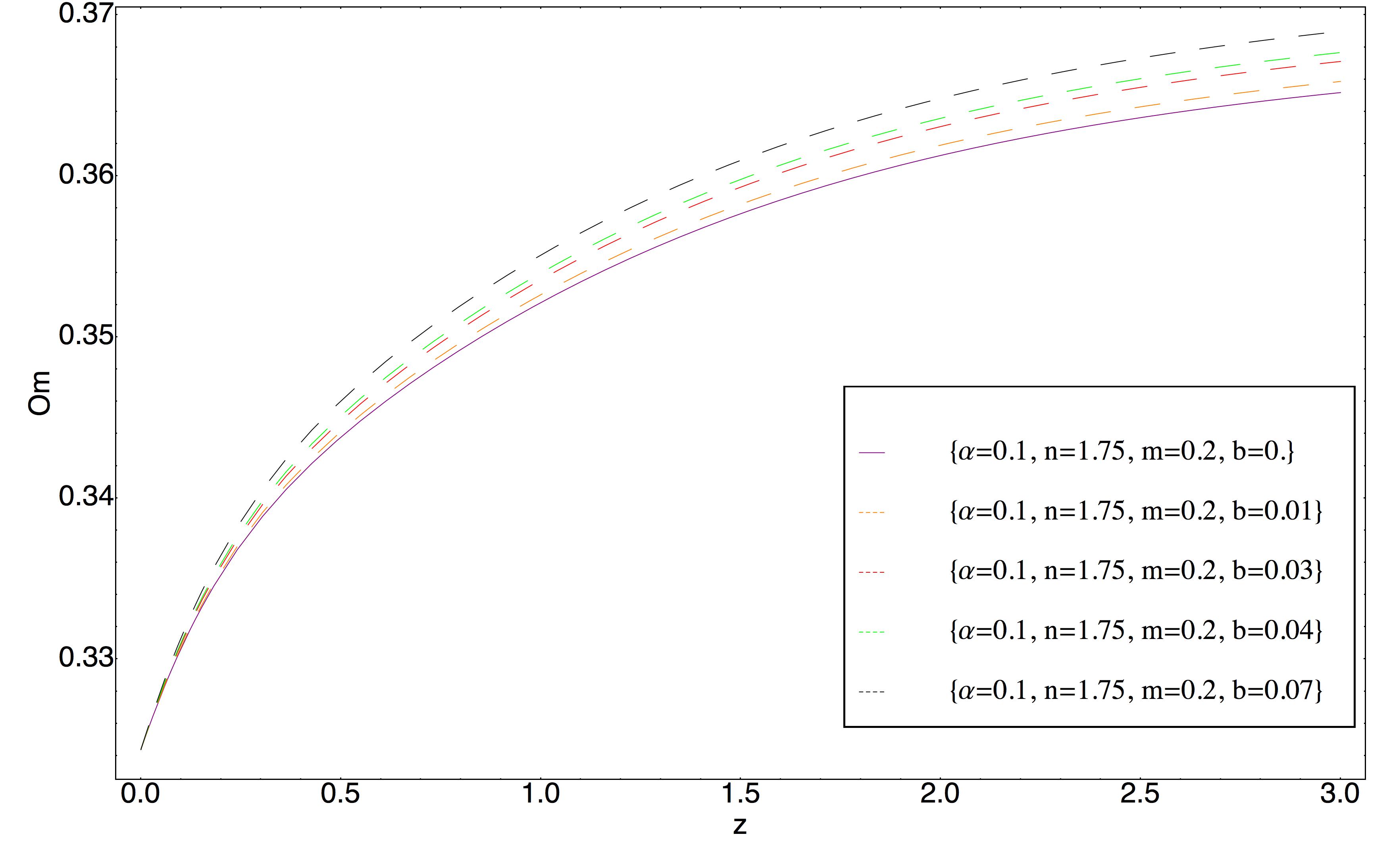}  &
\includegraphics[width=80 mm]{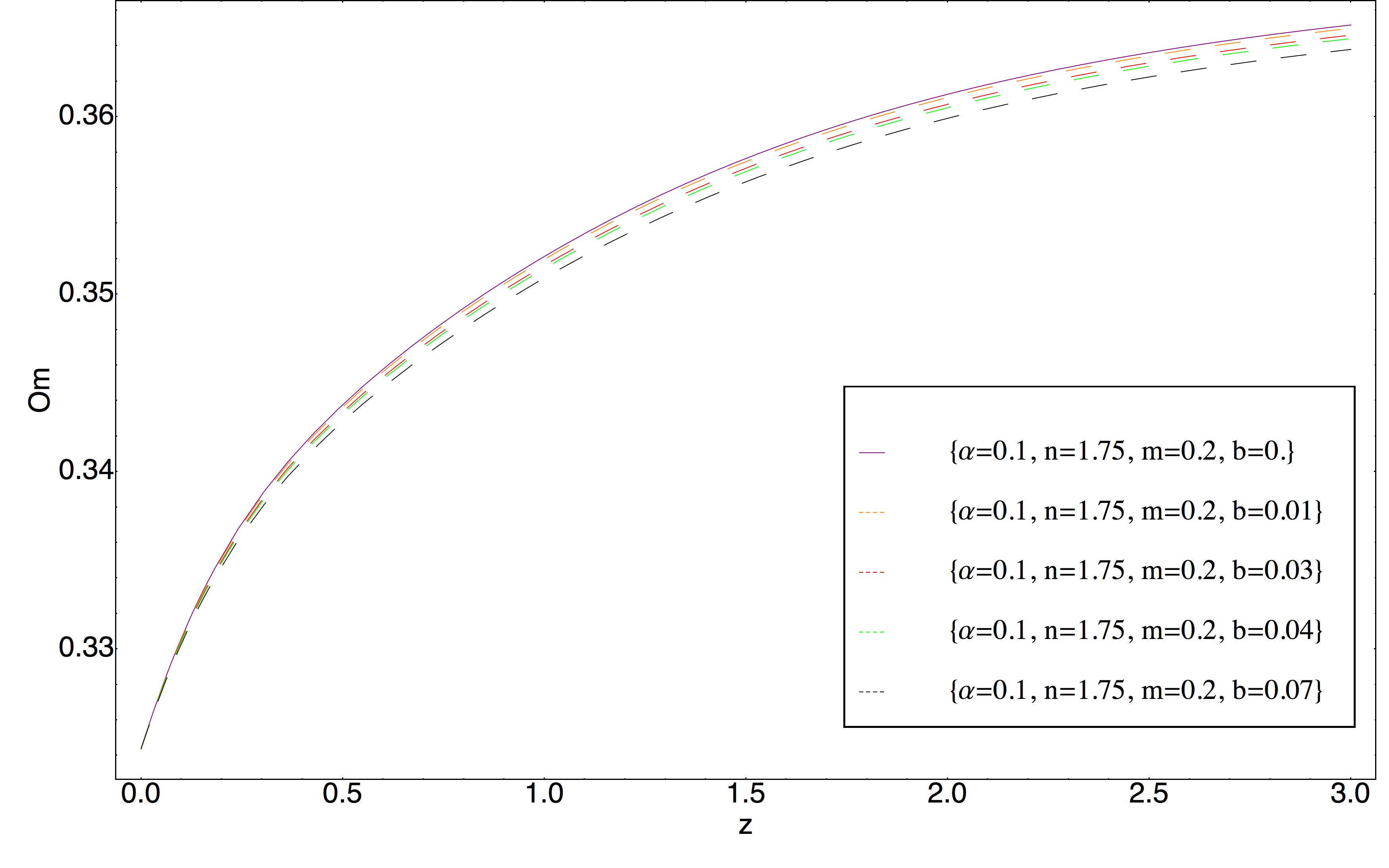}  &
 \end{array}$
 \end{center}
\caption{The graphical behavior of $Om$ parameter against redshift $z$. The left plot represents the graphical behavior of the parameter for the model with non-gravitational interaction given by Eq.~(\ref{eq:Q1}). The right plot represents the graphical behavior of the parameter for the model with non-gravitational interaction given by Eq.~(\ref{eq:Q2}).}
 \label{fig:Fig5}
\end{figure}

On the other hand, in the case of Type V~("$\omega$-singularity"), for $t \to t_{s}$, $a \to \infty$, $\rho_{eff} \to 0$ and $|P_{eff}| \to 0$ but the equation of state parameter $\omega \to \infty$~\cite{Sing1}. The presented list with $\omega$-singularity and the singularity introduced in Ref.~\cite{Sing2} are the current knowledge on spacetime singularities in the literature. The study of singularities allows to have an additional constraints providing a way for better understanding suggested cosmological models. On the other hand, in order to classify finite time future singularities we can use $Om$ analysis directly instead of studying the equation of state parameter. For instance, the graphical behavior of $Om$ parameter presented in Fig.~(\ref{fig:Fig4}) and Fig.~(\ref{fig:Fig5}), indicates that considered combination of dark components has quintessence nature and the background dynamics is such that the quintessence nature will be preserved from higher redshifts up to present epoch. 

However, for some values of the parameters of the models $Om$ parameter is a decresing function from redshift, indicating that in future a phantom behavior of the considered combination of dark components could be observed. On the other hand, this can be a signal about an existence of one of the singularities mentioned above. However, from the Fig.~(\ref{fig:Fig4}) we observe that it is also possible to have an increasing behavior for $Om$ parameter excluding a phantom nature of the model in near future. We see that for considered models $Om$ analysis demonstrates and indicates differences between considered models and $\Lambda$CDM standard model. Moreover, in this case $Om$ analysis is also useful to study induced differences caused by different values of the parameters of the models. The study of the non - interacting model shows that the parameter $\alpha$ can induce either Type II~(“The Sudden Singularity), or Type IV~("Generalized Sudden Singularity") singularities. In this case we observed that $\omega_{DE}$ and $\omega_{tot}$ of the effective fluid will have smooth and continuous behavior on $(-1.08,-0,98)$ range for $z \in [-1,0]$. On the other hand, for low values of parameter $m$ ($m \in (0,1)$) Type IV singularity could be formed in considered non-interacting model, while for the highest values of the same parameter the final fate of the Universe is Type II singularity. In summary, the study shows that characteristic singularities for non-interacting model are either Type II~(“The Sudden Singularity), or Type IV~("Generalized Sudden Singularity") depends on the values of the parameters. 

The consideration of non-gravitational interaction given by Eq.~(\ref{eq:Q1}) shows an interesting picture, in particular, we observed that for the higher values of parameter $b$ the background dynamics will tend recent Universe to Type V~("$\omega$-singularity") singularity. While for lower values of parameter $b \in [0,0.03]$ the background dynamics is such that the final fate of the Universe is either Type II singularity, or Type IV singularity depends on the values of the others parameters. This picture is observed for the values of the parameters allowing to obtain late time tachyonic cold dark matter. However, the consideration of non-gravitational sign-changeable interaction, Eq.~(\ref{eq:Q2}), determines the final fate of the Universe to be only Type IV singularity. As was expected considered non-gravitational interactions strongly affect on the singularity formation time $t_{s}$. Unfortunately, with a line of other research presented in literature, in this paper once again we convinced that form the present and higher redshift behavior of $Om$ parameter it is not possible to determine the type of finite time future singularities for a cosmological model. On the other hand, such possibility definitely will significantly increase the role of $Om$ analysis in modern cosmology.

\section{Tachyonic matter with specific $a(t)$ scale factor}\label{sec:SF}

The structure of considered cosmological model allows to determine the form of the potential and scalar field describing tachyonic dark matter, when explicit forms of the scale factor and non-gravitational interaction term are given. In this section $2$ forms of the scale factor will be considered to recover the behavior of the energy densities of dark components. To simplify the presentation of the results we organised appropriate subsections.

\subsection{Scale factor: $a = t^{n}$}
We start our study from the case when scale factor has the following form
\begin{equation}
a = t^{n}.
\end{equation} 
It is known that if $n>1$, then we have an accelerated expansion with
\begin{equation}
q = -1 + \frac{1}{n}.
\end{equation}
Let us assume that interaction between tachyonic dark matter and dark energy has the following form
\begin{equation}\label{eq:Q1_1}
Q = 3 H b \rho_{c} + \gamma \dot{\rho}_{c}, 
\end{equation}
then the dynamics of the energy density of Chaplygin gas will have the following form 
\begin{equation}
\rho_{c} = \left(\frac{3 (\alpha +1) A_{0} \left(t^n\right)^{-m}}{3 (\alpha +1)(1+b)-(\gamma +1) m}+ \rho_{c0} t^{-\frac{3 (\alpha +1) (b+1) n}{\gamma +1}}\right){}^{\frac{1}{\alpha +1}}. 
\end{equation}
On the other hand from the first Friedmann equation for the energy density of the tachyonic matter we will have 
\begin{equation}
\rho_{T} = \frac{3 n^2}{t^2}-\left(\frac{3 (\alpha +1) A_{0} \left(t^n\right)^{-m}}{3 (\alpha +1)(1+b)-(\gamma +1) m}+ \rho_{c0} t^{-\frac{3 (\alpha +1) (b+1) n}{\gamma +1}}\right)^{\frac{1}{\alpha +1}}.
\end{equation}
It is easy to see that when $t \to \infty $, then to have finite $\rho_{c}$ and $\rho_{T}$ the parameters of the model should satisfy to the following constraints: $nm > 0$ and $\frac{3 (\alpha +1) (b+1) n}{\gamma +1} > 0$. On the other hand with $\gamma = 0$ and $b = 0$ we can recover the cases corresponding to $Q = 3 H b \rho_{c}$ and $Q = \gamma \dot{\rho}_{c}$, respectively. On the other hand, the form of the potential and the scalar field $\phi$ can be found numerically.

\subsection{Scale factor: $a_{0} t^n+b_{0} \exp (k t)$}

If we consider a model of the Universe with the scale factor defined as follows

\begin{equation}
a = a_{0} t^n+b_{0} \exp (k t),
\end{equation} 
then for the dynamics of the energy density of the Chaplygin gas we will obtain
\begin{equation}
\rho_{c} = \left(\frac{3 (\alpha +1) A_{0} \left(a_{0} t^n+b_{0} e^{k t}\right)^{-m}}{3 (\alpha +1)(1+b)-(\gamma +1) m}+ \rho_{0c} \left(a_{0} t^n+b_{0} e^{k t}\right)^{-\frac{3 (\alpha +1) (b+1)}{\gamma +1}}\right){}^{\frac{1}{\alpha +1}},
\end{equation}
when the non-gravitational interaction is given by Eq.~(\ref{eq:Q1_1}). In this model the deceleration parameter reads as
\begin{equation}
q = -\frac{\left(a_{0} t^n+ b_{0} e^{k t}\right) \left(a_{0} (n-1) n t^n+b_{0} k^2 t^2 e^{k t}\right)}{\left(a_{0} n t^n+b_{0} k t e^{k t}\right)^2}.
\end{equation}
The form of the potential and the scalar field $\phi$ can be found numerically as in previous case. Moreover, with $\gamma = 0$ and $b = 0$ we can recover the cases corresponding to $Q = 3 H b \rho_{c}$ and $Q = \gamma \dot{\rho}_{c}$, respectively. Whilst $b_{0}=0$ will reduce the model to the model considered above. 

\section{Discussion}\label{sec:Disc}

In this paper we considered cosmological models representing dark energy and dark matter as a modified varying Chaplygin gas and tachyonic matter with a specific $V(\phi) = V_{0} \phi^{n}$ potential, respectively. The study shows that non - interacting model is free from the cosmological coincidence problem. Moreover, for appropriate values of the parameters tachyonic field will be a cold dark matter with $\omega_{DM} \approx 0$ at higher and lower redshifts. Moreover, the parameter $n$ plays the central role in the transition from dark energy to cold dark matter and an increase of $n$ will decrease the transition redshift. In case of non - interacting model an increase of $\alpha$ parameter describing varying Chaplygin gas gives a decrease in transition redshift and increases the present day value of the deceleration parameter $q$. On the other hand, the present day value of $\omega_{DM}$ does not affected strongly by the change of $\alpha$ parameter. Moreover, the analysis shows that a significant impact rised due to the parameter $\alpha$ on the behavior of EoS of the tachyonic field will be observed for $z \in [0.4,2]$ redshifts. In summary, we observed that it is possible to parametrize the energy budget of the Universe using a varying Chaplygin gas and a tachyonic matter with a specific potential to explain the accelerated expansion free from the cosmological coincidence problem. The study of the background dynamics involving two forms of non - gravitational interactions $Q=3 b H \rho_{de}$ and $Q=3 b q H \rho_{de}$ had been organised as well. In order to classify finite time future singularities we first used $Om$ analysis indicating that considered combination of dark components has quintessence nature and the background dynamics is such that the quintessence nature will be preserved from higher redshifts up to present epoch~(for the values of the parameters giving tachyonic cold dark matter). However, a decreasing nature of $Om$ parameter shows that there is a possibility of observing a phantom nature of the model in future. The study shows that in case of non-interacting model, the fate of the Universe will be either Type II~(“The Sudden Singularity), or Type IV~("Generalized Sudden Singularity") singularity. On the other hand, the consideration of $Q=3 b H \rho_{de}$ non-gravitational interaction in addition to Type II and Type IV singularities will induce also Type V~("$\omega$-singularity") singularity. Whilst in the model with $Q=3 b q H \rho_{de}$ non-gravitational interaction only Type IV singularity will be observed. Moreover, $2$ forms of the scale factor has been considered separately allowing analytically obtain the behavior of the cosmological parameters without specifying an explicit form of the potential for tachyonic matter~(the potential and field can be recovered numerically), when $Q = 3 H b \rho_{c} + \gamma \dot{\rho}_{c}$.

The considered model shows an existance of a new possibility to parametrize the dark side of the recent Universe. On the other hand, to have a better understanding future research is needed. In particular, other forms of the potential and non-gravitational interaction should be considered. Besides $Om$ analysis, $Omh^{2}$ can be used to obtain preliminary constraints on the parameters. Moreover, according to the discussion organised in this paper it is extremely important to classify future finite time singularities for new models and compare with the results obtained in this paper. A study on these questions with a construction of some modified theories of gravity~(for instance f(R) and f(T) theories of gravity) for the models will be discussed elsewhere shedding a light on proposed structure of the dark side of the Universe.


\begin{thebibliography}{1}

\bibitem{Riess}
A.G. Riess et al., Astron. J. 116 (1998) 1009; S. Perlmutter et al., Astrophys. J. 517 (1999) 565; R. Amanullah et al., Astrophys. J. 716 (2010) 712; A.C. Pope et al. Astrophys. J. 607 (2004) 655 [arXiv:0401249[astro-ph]];
D.N. Spergel et al. Astrophys. J. Supp. 148 (2003) 175 [arXiv:0302209[astro-ph]], Planck Collaboration, A $\&$ A 594, A13 (2016)


\bibitem{Steinhardt}
P.J. Steinhardt, "Critical Problems in Physics" (1997), Prinston University Press; M. Roos, John Wiley $\&$ Sons, Ltd, ISBN: 978-1-118-92332-0 (2015); J. Sola and H. Stefancic, Phys. Lett. B 624 (2005) 147; I.L. Shapiro and J. Sola, Phys. Lett. B 682 (2009) 105; B. Ratra and P. J. E. Peebles, Phys. Rev. D 37 (1988) 3406; C. Wetterich, Nucl. Phys. B 302 (1988) 668; I. Zlatev, L. M. Wang and P. J. Steinhardt, Phys. Rev. Lett. 82 (1999) 896; Z. K. Guo, N. Ohta and Y. Z. Zhang, Mod. Phys. Lett. A 22 (2007) 883; S. Dutta, E. N. Saridakis and R. J. Scherrer, Phys. Rev. D 79 (2009) 103005; E.N. Saridakis and S. V. Sushkov, Phys. Rev. D 81 (2010) 083510; R. R. Caldwell, M. Kamionkowski and N.N. Weinberg, Phys. Rev. Lett. 91 (2003) 071301; R. R. Caldwell, Phys. Lett. B 545 (2002) 23; J. Yoo and Y. Watanabe, Int. J. Mod. Phys. D 21 (2012) 1230002; S. Nojiri and S. D. Odintsov, Phys. Lett. B 562 (2003) 147 [arXiv:0303117[hep-th]]; P. Singh, M. Sami and N. Dadhich, Phys. Rev. D 68 (2003) 023522; J.M. Cline, S. Jeon and G.D. Moore, Phys. Rev. D 70 (2004) 043543; V.K. Onemli and R.P. Woodard, Phys. Rev. D 70 (2004) 107301; W. Hu, Phys. Rev. D 71 (2005) 047301; M.R. Setare and E. N. Saridakis, JCAP 0903 (2009) 002; E.N. Saridakis, Nucl. Phys. B 819 (2009) 116; B. Feng, X. L. Wang and X. M. Zhang, Phys. Lett. B 607 (2005) 35; E. Elizalde, S. Nojiri and S.D. Odintsov, Phys. Rev. D 70 (2004) 043539 [arXiv:0405034[hep-th/]]; A. Kamenshchik, U. Moschella and V. Pasquier, Phys. Lett. B511 (2001) 265; N.Bilic, G.B. Tupper and R.D. Viollier, Phys. Lett. B535 (2002) 17; M.C. Bento, O. Bertolami and A.A. Sen, Phys. Rev. D67 (2003) 063003


\bibitem{Guo}
Z.K. Guo, et al., Phys. Lett. B 608 (2005) 177; M.-Z Li, B. Feng, X.-M Zhang, JCAP, 0512 (2005) 002; B. Feng, M. Li, Y.-S. Piao and X. Zhang, Phys. Lett. B 634 (2006) 101; S. Capozziello, S. Nojiri and S.D. Odintsov, Phys. Lett. B 632 (2006) 597 [arXiv:0507182[hep-th]]; W. Zhao and Y. Zhang, Phys. Rev. D 73 (2006) 123509; Y.F. Cai et al., JHEP 0710 (2007) 071; E.N. Saridakis and J.M. Weller, Phys. Rev. D 81 (2010) 123523; Y.F. Cai et al., JCAP 0803 (2008) 013; M.R. Setare and E.N. Saridakis, Phys. Lett. B 668 (2008) 177; M.R. Setare and E.N. Saridakis, Int. J. Mod. Phys. D 18 (2009) 549; Y. F. Cai et al., Phys. Rept. 493 (2010) 1; T. Qiu, Mod. Phys. Lett. A 25 (2010) 909; S.D.H. Hsu, Phys. Lett. B 594 (2004) 13; M. Li, Phys. Lett. B 603 (2004) 1; Q.G. Huang and M. Li, JCAP 0408 (2004) 013; M. Ito, Europhys. Lett. 71 (2005) 712; X. Zhang and F.Q. Wu, Phys. Rev. D 72 (2005) 043524; D. Pavon and W. Zimdahl, Phys. Lett. B 628 (2005) 206; L.P. Chimento and A.S. Jakubi, Int. J. Mod. Phys. D5 (1996) 71; T. Matos, F.S. Guzman and L.A. Urena-Lopez, Class. Quant. Grav. 17 (2000) 1707; L.A. Urena-Lopez and T. Matos, Phys. Rev. D62 (2000) 081302; A. Gonzalez, T. Matos amd L. Quiros, Phys. Rev. D71 (2005) 084029


\bibitem{Nojiri}
S. Nojiri and S.D. Odintsov, Gen. Rel. Grav. 38 (2006) 1285 [arXiv:0506212[hep-th]]; E. Elizalde, S. Nojiri, S.D. Odintsov and P. Wang, Phys. Rev. D 71 (2005) 103504 [	arXiv:0502082[hep-th]]; H. Li, Z.K. Guo and Y.Z. Zhang, Int. J. Mod. Phys. D 15 (2006) 869; E.N. Saridakis, Phys. Lett. B 660 (2008) 138; E.N. Saridakis, JCAP 0804 (2008) 020
E.N. Saridakis, Phys. Lett. B 661 (2008) 335; R.G. Cai, Phys. Lett. B 657 (2007) 228; H. Wei and R.G. Cai, Phys. Lett. B 660 (2008) 113; H. Wei and R.G. Cai, Eur. Phys. J. C 59 (2009) 99; I. Brevik et al, arXiv:1706.02543; G.M. Kremer, Phys. Rev. D 68 (2003) 123507; W. Chakraborty and U. Debnath, Astropys Space Sci (2008) 315:73-78


\bibitem{Hao}
W. Hao, Common. Theory. Phys. 56 (2011) 972-980; H.Wei, Nucl. Phys. B 845 (2011) 381; A.P. Billyard and A.A. Coley, Phys. Rev. D. 61 (2000) 083503 [arXiv:9908224[astro-ph]]; J. Sadeghi et al, JCAP 12, 031 (2013) [arXiv:1308.3450 [gr-qc]]; M Khurshudyan, Eur. Phys. J. Plus 131: 25 (2016); M. Khurshudyan, Mod. Phys. Lett. A, 31, 1650055 (2016); M. Khurshudyan, Mod. Phys. Lett. A, 31, 1650097 (2016); M. Khurshudyan, Symmetry, 8(11), 110 (2016); M. Khurshudyan, R Myrzakulov, European Physical Journal C 77 (2), 65 (2017) [arXiv:1509.02263 [gr-qc]]; M. Zh. Khurshudyan and A. N. Makarenko, Astrophys Space Sci 361: 187 (2016) [arXiv:1606.06590 [gr-qc]]; M. Khurshudyan et al, Astrophys Space Sci 357: 113 (2015) [arXiv:1307.7859 [gr-qc]]; M. Khurshudyan, Astrophys Space Sci 360: 33 (2015) [arXiv:1510.07962 [physics.gen-ph]]; M. Khurshudyan, Astrophys Space Sci 361: 232 (2016) [	arXiv:1606.05264 [gr-qc]]; M. Khurshudyan, As. Khurshudyan [arXiv:1708.02293 [gr-qc]]; M. Khurshudyan, M. Hakobyan, As. Khurshudyan [arXiv:1707.02856 [gr-qc]]; L. P. Chimento, Phys. Rev. D 81, 043525 (2010); L. P. Chimento, Phys. Rev. D69, 123517 (2004); M. Khurshudyan [arXiv:1301.1021v2 [gr-qc]]; Xi-ming Chen, et al., Int.J.Theor.Phys. 53 (2014) 469-481 [arxiv:1111.6743 [astro-ph.CO]]; Xi-ming Chen et al., JCAP 0904 (2009) 001 [arXiv:0812.1117 [gr-qc]]; V. Faraoni et al., Phys.Rev. D90 (2014) no.6, 063510 [arXiv:1405.7288 [gr-qc]]


\bibitem{MGR}
S. Nojiri and S. D. Odintsov, Int. J. Geom. Methods Mod. Phys. 04, 115 (2007) [arXiv:0601213[hep-th]];
S. Nojiri and S. D. Odintsov, Phys. Rept. 505:59-144 (2011) [arXiv:1011.0544 [gr-qc]];
S. Nojiri and S. D. Odintsov, Phys. Rev. D 68, 123512 (2003) [arXiv:0307288 [hep-th]];
S. Nojiri and S. D. Odintsov, Phys. Rev. D 74, 086005 (2006) [arXiv:0608008 [hep-th/]];
S. Capozziello et al, Phys. Lett. B 639:135-143 (2006) [arXiv:0604431 [astro-ph]];
G. Cognola et al, Phys. Rev. D 77, 046009 (2008) [arXiv:0712.4017 [hep-th]];
Yi-Fu Cai et al, Rept. Prog. Phys. 79, no. 4, 106901 (2016) [arXiv:1511.07586 [gr-qc]];
J. B. Dent et al, JCAP 1101 (2011) 009 [arXiv:1010.2215 [astro-ph.CO]];
S. Nesseris et al, Phys. Rev. D 88, 103010 (2013) [arXiv:1308.6142 [astro-ph.CO]];
V.K. Oikonomou, E. N. Saridakis, Phys. Rev. D 94, 124005 (2016);
T. Clifton et al, Physics Reports 513, 1-189 (2012);
S. Capozziello, M. De Laurentis, Physics Reports 509, 167321 (2011);
K. Bamba, S. D. Odintsov, Symmetry 2015, 7, 220-240 [arXiv:1503.00442 [hep-th]];
S. Nojiri et al [arXiv:1705.11098]; S. Chattopadhyay et al, Eur. Phys. J. C 74:3080 (2014) [arXiv:1401.8208 [gr-qc]]

\bibitem{PCR}
J.A.S.Lima., D.Singleton, Physics Letters B 762, 506511 (2016); J. F. Jesus, S. H. Pereira, JCAP 07, 040 (2014); A. Paliathanasis et al, Phys. Rev. D 95, 103516 (2017); J. Chen et al. Eur. Phys. J. C 72: 1861 (2012); R. C. Nunes, S. Pan, Mon. Not. Roy. Astron. Soc, 459, 673-682 (2016)


\bibitem{Zong}
Zong-Kuan Guo and Yuan-Zhong Zhang, Phys. Lett. B 645:326-329 (2007) [arXiv:0506091[astro-ph]]


\bibitem{Sen}
A. Sen, JHEP 0204, 048 (2002) [arXiv:0203211 [hep-th]]; A. Sen, JHEP 0207, 065 (2002) [arXiv:0203265 [hep-th]]; A. Sen, Mod.Phys.Lett.A17:1797-1804 (2002) [arXiv:0204143 [hep-th]]

\bibitem{Murli}
Murli Manohar Verma and Shankar Dayal Pathak, A Tachyonic Scalar Field with Mutually Interacting Components, Int J Theor Phys, DOI 10.1007/s10773-012-1116-8.

\bibitem{Om}
V. Sahni et al., Phys. Rev. D 78 103502 (2008)


\bibitem{Sing}
S. Nojiri, S. D. Odintsov, S. Tsujikawa, Phys. Rev. D 71:063004 (2005) [arXiv:hep-th/0501025]

\bibitem{Sing1}
M. P. Dabrowski and T. Denkieiwcz, Phys. Rev. D 79, 063521 (2009); M. P. Dabrowski and T. Denkiewicz, AIP Conf. Proc. 1241, 561 (2010); L. Fernandez-Jambrina, Phys. Lett. B 656, 9 (2007)

\bibitem{Sing2}
J. B. Jimenez et al. [arXiv:1607.06389 [gr-qc]]

\end{thebibliography}
\end{document}